\documentclass{article}
\usepackage{spconf,amsmath,graphicx}
\usepackage{algorithmic}
\usepackage{amssymb}
\usepackage[ruled,linesnumbered]{algorithm2e}
\usepackage{booktabs}
\usepackage{tabularx}
\usepackage{multicol}
\usepackage{multirow}
\usepackage{amsmath}
\usepackage{adjustbox}
\usepackage{graphicx}
\usepackage{subfigure}
\newcounter{mysubfigure} 

\usepackage{bbding}
\usepackage{cite}
\usepackage{enumitem} 
\usepackage{pifont}
\usepackage{threeparttable}
\RequirePackage{bibspacing}

\setlist[enumerate]{ 
  nosep,
  align=left, 
  leftmargin=*, 
}

\title{SSHNN: Semi-Supervised Hybrid NAS Network for Echocardiographic Image Segmentation}
\name{Renqi Chen\textsuperscript{\textnormal{12}}, Jingjing Luo\textsuperscript{\textnormal{12*}}\thanks{*Corresponding authors. Email: \{luojingjing, yzk\}@fudan.edu.cn. This research was funded by the National Key Research and Development Program of China (2022YFC361400/2022YFC3601401) and the Natural Science Foundation of China (U1913216).}, Fan Nian\textsuperscript{\textnormal{1}}, Yuhui Cen\textsuperscript{\textnormal{1}}, Yiheng Peng\textsuperscript{\textnormal{1}}, and Zekuan Yu\textsuperscript{\textnormal{1*}}}
\address{\textsuperscript{1}Institute of AI\&Robotics, Academy for Engineering\&Technology, Fudan University, Shanghai, China\\
\textsuperscript{2}Engineering Research Center of AI and Robotics, Ministry of Education, Shanghai, China}
\begin{document}
%
\maketitle
\begin{abstract}
Accurate medical image segmentation especially for echocardiographic images with unmissable noise requires elaborate network design. Compared with manual design, Neural Architecture Search (NAS) realizes better segmentation results due to larger search space and automatic optimization, but most of the existing methods are weak in layer-wise feature aggregation and adopt a ``strong encoder, weak decoder" structure, insufficient to handle global relationships and local details. To resolve these issues, we propose a novel semi-supervised hybrid NAS network for accurate medical image segmentation termed SSHNN. In SSHNN, we creatively use convolution operation in layer-wise feature fusion instead of normalized scalars to avoid losing details, making NAS a stronger encoder. Moreover, Transformers are introduced for the compensation of global context and U-shaped decoder is designed to efficiently connect global context with local features. Specifically, we implement a semi-supervised algorithm Mean-Teacher to overcome the limited volume problem of labeled medical image dataset. Extensive experiments on CAMUS echocardiography dataset demonstrate that SSHNN outperforms state-of-the-art approaches and realizes accurate segmentation. Code will be made publicly available.  
\end{abstract}
\vspace{0mm}
\section{Introduction}\label{sec:intro}
\vspace{-3mm}
2D echocardiography is a critical medical imaging technique for clinical routine to measure the cardiac morphology and function, and further obtaining a diagnosis \cite{leclerc2019deep}. Due to noise, artifacts and low contrast caused by uneven illumination, organ boundaries in echocardiographic images are blurred \cite{li2023echoefnet}. Besides, structures in echocardiographic images appear differently at various scales, requiring the model effectively handles the multi-scale features for medical image segmentation. 

Benefiting from receptive field, Convolutional Neural Networks (CNNs) represented extraordinary feature extraction ability and are frequently adopted \cite{girshick2015fast, chen2017deeplab, zhao2017pyramid}. Especially in medical area, the proposal of U-shaped encoder-decoder network has propelled semantic segmentation to the superior development \cite{ronneberger2015u, zhou2018unet++, xiao2018weighted, guan2019fully}. Moreover, when the Transformer \cite{vaswani2017attention} is introduced into computer vision and converted to vision Transformer (ViT)\cite{dosovitskiy2020image} to enhance the global receptive ability, the combination of U-shaped network with Transformer is focused, such as TransUNet \cite{chen2021transunet}, Swin Transformer \cite{liu2021swin}, etc. The ever-proposed manual designed networks improve segmentation accuracy while becoming more complex. 

To resolve the difficulty and complexity of network designing, Neural Architecture Search (NAS) is proposed, aiming to design automatically and accurately. Among them, NAS-UNet \cite{weng2019unet} replaces the CNN portion with searchable cells to reduce network parameters. In order to expand search space for better generalization, Hierarchical NAS (HNAS) also searches in layer-level, such as Auto-DeepLab \cite{liu2019auto}, Dynamic routing \cite{li2020learning}, etc. Moreover, HCT-Net \cite{yu2023hct} adopts ViT to add context information. However, these methods employ normalized scalars for feature fusion in lay-wise optimizing, prone to local detail loss caused by insufficient parameters. Besides, despite having usable multi-scale features, decoder only depends on concatenation and convolution layers, contributing to ``strong encoder, weak decoder", incapable to capture sufficient context and local information \cite{liu2019auto}. Hence, our focus is enhance model segmentation ability through the integration of exemplary manual design and HNAS.

\begin{figure*}[ht]
    \centering
    \includegraphics[width=0.9\linewidth]{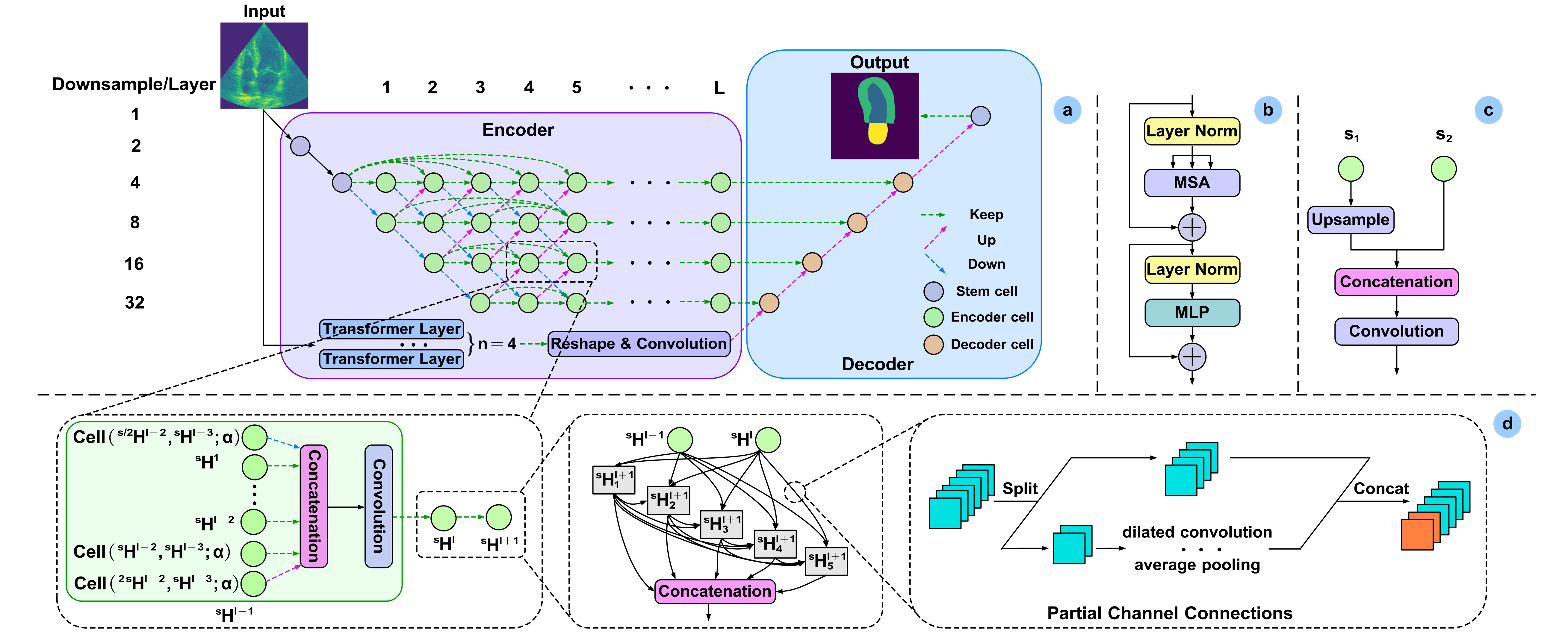}
    \vspace{-5mm}
    \caption{The network framework of SSHNN. (a) Overview of SSHNN. Dashed lines denote candidate connections. We only show several connections among all skip connections to keep clarity. (b) Transformer layer module. (c) Decoder cell module. (d) Encoder cell module. \textit{Left}: Feature fusion details. \textit{Middle}: Inner cell search space. \textit{Right}: Partial channel connections.}\label{Structure}
\end{figure*}

In this paper, we propose a novel hybrid NAS with semi-supervised learning considering the size limitation of labeled medical image dataset, named SSHNN. In the design of HNAS, we replace simple scalars with convolutions for local feature aggregation, pursuing higher degree of flexibility. To overcome ``strong encoder, weak decoder", add U-shaped decoder after HNAS to recover multi-scale features into original size, and Transformer is utilized to compensate for global context. Experiments prove SSHNN outperforms state-of-the-art (SOTA) methods on public medical image dataset on echocardiography: CAMUS \cite{leclerc2019deep}.
\vspace{-6mm}
\section{Methodologies}
\vspace{-4mm}
In this section, We elaborate SSHNN in four parts: HNAS design, decoder, semi-supervised learning and optimization strategy. The network framework is shown in Fig. \ref{Structure}. 
\vspace{-5mm}
\subsection{HNAS design}
\vspace{-2mm}
\noindent \textbf{Inner Cell:} Cell search space can be represented by a directed acyclic graph, consisting of blocks and edges, denoting the mapping from input tensors to 1 output tensor, and candidate operations, respectively. For the $i^{th}$ block of a cell in the $l^{th}$ layer, define a tuple $(I_{i}^{l}, O_{i}^{l})$ to denote the mapping, where $I_{i}^{l} \in\mathcal{I}$ denotes the input tensor, 
and $O_{i}^{l}\in\mathcal{O}$ denotes the candidate operation. The set of input tensors consists of previous cell's output $H^{l-1}$, previous-previous cell's output $H^{l-2}$, and previous blocks' output in the current cell ${H_{1}^{l},\dots,H_{i-1}^{l}}$. The set of candidate operations $\mathcal{O}$ includes: depthwise-separable convolution, dilated convolution, average polling, max polling, skip connection and no connection. 

To reduce memory cost, partial channel connections \cite{xu2019pc} are adopted, where $1/n$ portion of dimensional features are sent to the cell and the rest features remain unchanged. Moreover, continuous relaxation \cite{liu2018darts} is reused as differentiable search space is pursued and then the stochastic gradient descent (SGD) can be applied. Thus, the output tensor of block:
\vspace{-3mm}
\begin{equation}
\resizebox{0.9\hsize}{!}{$
\begin{aligned}
    H_{i}^{l} &= \sum_{H_{j}^{l}\in\mathcal{I}}\sum_{O_{k}^{l}\in \mathcal{O}}\frac{exp\{\alpha_{j\rightarrow i}^{k}\}}{\sum_{m=1}^{|\mathcal{O}|}exp\{\alpha_{j\rightarrow i}^{m}\}}\cdot O_{k}^{l}(Part_{j\rightarrow i}\circ H_{j}^{l})\\
  & +(1-Part_{j\rightarrow i})\circ H_{j}^{l}
\end{aligned}$}
\end{equation}
where $Part_{j\rightarrow i}$ is sampling mask for channel selection,
and $\alpha$ are normalized scalars, called architecture parameters, measuring the weight of candidate operation. To ensure the differentiable 
back propagation, softmax is implemented. Finally, the output tensor $H^{l} = Concat(\{H_{i}^{l}|i<=B\})$, where $Concat(\cdot)$ is concatenation and $B$ is the number of blocks. Cell level search is denoted as $H^{l}=Cell(H^{l-1},H^{l-2};\alpha)$.\\
\noindent\textbf{Out Layer:} Layer level search aims at finding the optimal network backbone within network search space for specific dataset, with the specific procedure is combine features from different resolutions for better feature extraction. In past researches \cite{liu2019auto,liu2021mixsearch,zhang2021dcnas}, normal algorithm uses another architecture parameter $\beta$ for feature fusion by linear combination. However, the amount of $\beta$ is a not large, which is usually in the hundreds, but are responsible for complex feature fusion, especially after applying partial channel connection. By estimation in Sec. \ref{ablation study}, linear combination cannot fully fuse multi-scale features since hundreds of channels represent different features, which needs more parameters to elaborate. 

Thus, we replace normalized scalars with convolution kernels to realize higher degree of flexibility, shown in Fig. \ref{Structure}(d, $Left$). Furthermore, long and short skip connection is utilized due to an incredible ability in image segmentation \cite{zhou2018unet++, zhang2021dcnas}. The whole representation of layer design:
\vspace{-2mm}
\begin{equation}
\resizebox{0.95\hsize}{!}{$
\begin{aligned}
  ^{s}H^{l} = & \ Conv(Concat(Down(Cell(^{\frac{s}{2}}H^{l-1},^{s}H^{l-2};\alpha)),\\
  &Up(Cell(^{2s}H^{l-1},^{s}H^{l-2};\alpha)),Cell(^{s}H^{l-1},^{s}H^{l-2};\alpha),\\
  &\{^{s}H^{l'}\in{^{s}H}|l'<l\}))
\end{aligned}$}
\end{equation}
where $Conv(\cdot)$ denotes the convolution, transforming fused features to the same channel counts of $s$-resolution, $Down(\cdot)$ denotes downsampling, and $Up(\cdot)$ denotes upsampling.
\vspace{-5mm}
\subsection{Decoder}
\vspace{-3mm}
Rather than simply concatenating multi-scale features and then processing by convolution layers to revert features to the original image size after searching optimal network structure, where these layers act as decoder and HNAS acts as encoder essentially, we apply ViT to add global context and use U-shape decoder structure, shown in Fig. \ref{Structure}(a). 

Define input image $\mathbf{x}\in\mathbb{R}^{H\times W\times C}$, where $H\times W$ denotes spatial resolution and $C$ denotes channel counts. 
First, perform tokenization \cite{yan2020ms}. Reshape $\mathbf{x}$ into $\{\mathbf{x}_{p}^{i}\in\mathbb{R}^{P\times P\times C}|i=1,\dots,N\}$ by $P\times P$ convolution (stride = $P$), where $N=\frac{HW}{P^{2}}$. 
Second, patch embedding and transformer. Map the patches $\mathbf{x}_{p}$ into a D-dimensional embedding space and add specific position embedding to keep position information, and then apply Transformers, including Multi-head Self-Attention (MSA) and Multi-layer Perceptron (MLP) blocks:
\vspace{-2mm}
  \begin{align}
  \small
    \mathbf{z}_{l}^{*} &= MSA(LN(\mathbf{z}_{l-1}))+\mathbf{z}_{l-1}\\
    \mathbf{z}_{l} &= MLP(LN(\mathbf{z}_{l}^{*}))+\mathbf{z}_{l}^{*}
  \end{align}
where $LN(\cdot)$ represents the layer normalization. 

Thus, Transformer output is $\mathbf{z}_{l}\in\mathbb{R}^{\frac{HW}{P^{2}}\times D}$ and then we reshape it into $\mathbb{R}^{\frac{H}{P}\times\frac{W}{P}\times D}$ for decoder. Herein, we need to combine the global features extracted from Transformers with the local features extracted from NAS. To avoid losing low-level or high-level details, use U-shape decoder is necessary.

We first use a 2D convolution to adjust the channels of encoded features from Transformer to have the same number of channels as $^{s}H^{l}$ for 
feature aggregation. Since we do not expect that transformers bring excessive parameters, $\frac{H}{P}$ and $\frac{W}{P}$ are small and then should be upsampled to $s$-resolution spatial size. After reshaping and resamping, concatenate two parts of features, followed with a convolution layer to match the number of channels with the next resolution $^{\frac{s}{2}}H^{l}$ channel counts, and then attain the output of current resolution. 

Similarly, ‘‘Upsample-Concatenation-Convolution" (shown in Fig. \ref{Structure}(c)) is implemented again for the next resolution features until combining $s=4$ resolution features. For segmentation purposes, finally a upsample layer and a convolution layer are used to recover the features to the full resolution and specific number of classes for predicting the dense output.
\vspace{-5mm}
\subsection{Semi-Supervised Learning}
\vspace{-3mm}
As medical image dataset always has a small amount of labeled images and a large amount of unlabeled data. Thus, we try to combine NAS with Mean Teacher method \cite{tarvainen2017mean}, to enhance the effectiveness and generalisation of the model.

For Mean Teacher, we have two models: $Student$ $f(\theta_{s})$ and $Teacher$ $f(\theta_{t})$, where $\theta_{s}$ and $\theta_{t}$ denote the network parameters of $Student$ and $Teacher$, respectively. $\theta_{t}$ are updated by the exponential moving average (EMA) of $\theta_{s}$:
\vspace{-2mm}
\begin{equation}
  \theta_{t,i}=\alpha\theta_{s,i-1}+(1-\alpha)\theta_{s,i}
\end{equation}
where $\alpha$ is a hyperparameter controlling the updating speed, $i$ is iteration times. Given labeled dataset $\mathcal{D}_{l}$ and unlabeled dataset $\mathcal{D}_{u}$, define the consistency regularization to update $\theta_{s}$:
\vspace{-2mm}
\begin{equation}
\resizebox{0.98\hsize}{!}{$
\begin{aligned}
  \mathcal{L}_{c} = \frac{\sum_{\mathcal{D}_{l}}{l}_{MSE}(S(p_{l}^{t}),S(p_{l}^{s}))}{m} + \frac{\sum_{\mathcal{D}_{u}}{l}_{MSE}(S(p_{u}^{t}),S(p_{u}^{s}))}{n}
\end{aligned}$}
\end{equation}
where $p_{l}^{t}$ and $p_{l}^{s}$ are the outputs of $Teacher$ and $Student$ on labeled images, $p_{u}^{t}$ and $p_{u}^{s}$ are the outputs on unlabeled images, $l_{MSE}$ is the Mean-Square Error, and $S(\cdot)$ is softmax function applied in channel dimension for scale controlling.

The supervised loss of $Student$ is computed by the output of $Teacher$ and the ground truth of labeled images:
\vspace{-2mm}
\begin{equation}
  \mathcal{L}_{s} = \frac{1}{m}\sum_{\mathcal{D}_{l}}l_{CE}(p_{l}^{t},y_{l})
\end{equation} 
where $l_{CE}$ is cross entropy loss function. Thus, the total loss used to train the $\theta_{s}$ is the sum of the consistency regularization and the supervised loss: $\mathcal{L}_{total}=\lambda_{0}\mathcal{L}_{s}+\lambda_{1}\mathcal{L}_{c}$, where $\lambda_{0}=1$ and $\lambda_{1}$ follows exponential ramp-up function in \cite{laine2016temporal}.
\vspace{-5mm}
\subsection{Optimization}
\vspace{-3mm}
To construct a differentiable computation graph, we use continuous relaxation for architecture parameters $\alpha$ controlling the connection in inner cell and differential convolution layers for architecture parameters $\gamma$ controlling the connection in outer layer, which makes gradient descent possible. In the training, we split the labeled and unlabeled training dataset into two disjoint sets respectively: $\mathcal{D}_{l,A}$ and $\mathcal{D}_{l,B}$, $\mathcal{D}_{u,A}$ and $\mathcal{D}_{u,B}$. The optimization in each epoch can be summarized as:
\begin{enumerate}
  \itemsep0em 
  \topsep0em 
  \parsep0em 
  \item Update weight parameters $w$ of $f(\theta_{s})$ by\\ $\nabla_{w}\mathcal{L}_{\mathcal{D}_{l,A},\mathcal{D}_{u,A}}(w,\alpha,\gamma)$ using $\mathcal{D}_{l,A}$ and $\mathcal{D}_{u,A}$
  \item Update weight parameters $w$ of $f({\theta_{t}})$ by EMA
  \item Update architecture parameters $\alpha$ and $\gamma$ of $f(\theta_{s})$ by $\nabla_{\alpha,\gamma}\mathcal{L}_{\mathcal{D}_{l,B},\mathcal{D}_{u,B}}(w,\alpha,\gamma)$ using $\mathcal{D}_{l,B}$ and $\mathcal{D}_{u,B}$
  \item Update architecture parameters $\alpha$ and $\gamma$ of $f(\theta_{t})$ by EMA
\end{enumerate}
where $\mathcal{L}$ is the total loss on the segmentation mini-batch.

\begin{figure*}
    \centering
    \includegraphics[width=\linewidth]{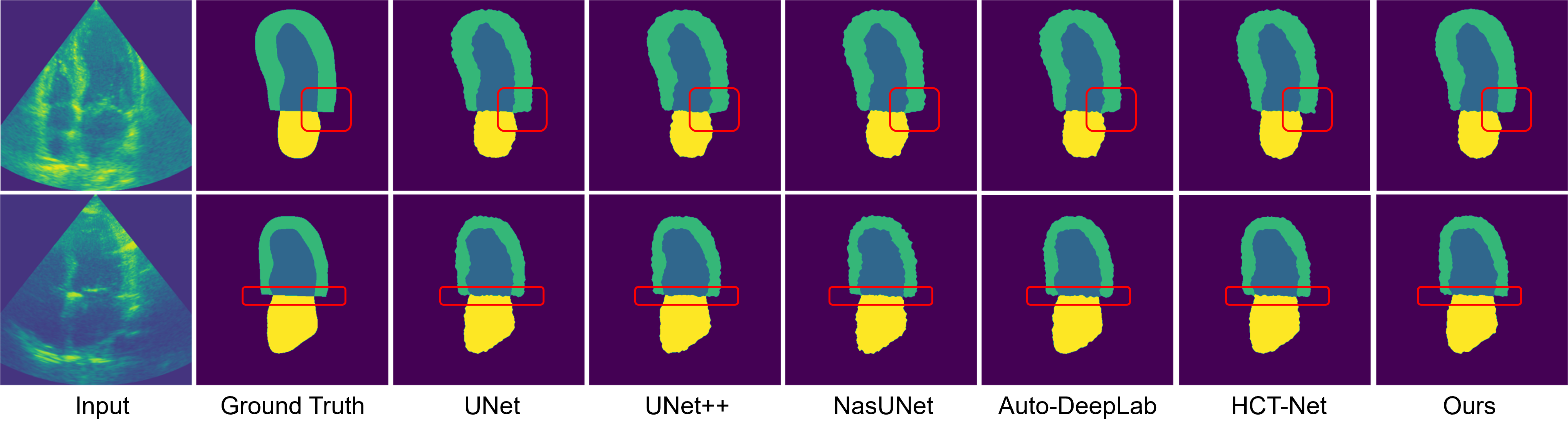}
    \vspace*{-11mm}
    \caption{The visual comparison of the proposed method with partial SOTA approaches on the CAMUS dataset.}\label{figure2}
\end{figure*}

\begin{table}[t]
    \caption{Quantitative comparison with SOTA methods based on manual network, pure NAS and hybrid NAS on CAMUS dataset. The best results are highlighted in bold.}\label{tableofexperiemnts}
  \centering
  \begin{adjustbox}{scale=0.78}  
  \begin{threeparttable}
  \begin{tabular}{ccccc}
  \toprule\hline
  \multirow{1}{*}{Type}      & Model & Params& Dice$\uparrow$  & IoU$\uparrow$       \\ \hline
  \multirow{5}{*}{Manual}    & {UNet \cite{ronneberger2015u}}                   &        {7.8M}  & {0.921$\pm$0.002}  &  0.859$\pm$0.003 \\  
     & UNet++ \cite{zhou2018unet++}     &    9.1M                     & {0.919$\pm$0.006}    &   0.855$\pm$0.009   \\  
         & Deeplabv3+ \cite{chen2018encoder}              &  {43.5M}     & {0.909$\pm$0.006}     &  0.839$\pm$0.009    \\  
         & Transfuse \cite{zhang2021transfuse}              &          {26.3M}                   & {0.923$\pm$0.004}    &   0.861$\pm$0.006   \\  
        & {PSPNet \cite{zhao2017pyramid}}   &  {49.0M}                    & {0.916$\pm$0.005}      &   0.850$\pm$0.008   \\ \hline
  \multirow{3}{*}{Pure NAS} & {NasUNet \cite{weng2019unet}}                &  {\textbf{0.8M}} & {0.914$\pm$0.004} &  0.845$\pm$0.005 \\ 
    & {Auto-DeepLab \cite{liu2019auto}}  &     {44.4M}                      & {0.918$\pm$0.003}     &  0.851$\pm$0.006    \\ 
    & {MixSearch \cite{liu2021mixsearch}}   &          {10.4M}                   & {0.917$\pm$0.004}      &   0.852$\pm$0.004   \\\hline  
    \multirow{4}{*}{Hybrid NAS}     & {HCT-Net \cite{yu2023hct}}&         {31.1M}   & {0.919$\pm$0.005}   &    0.856$\pm$0.005  \\ 
    & {SSHNN-S}          &     {13.6M}                    & {0.916$\pm$0.004}     &  0.850$\pm$0.006   \\
    & {SSHNN-M}          &        {23.5M}                    & 0.925$\pm$0.002   &  0.865$\pm$0.003   \\
    & {SSHNN-L}  &  {38.8M}  & \textbf{0.932$\pm$0.002}   &  \textbf{0.873$\pm$0.002}   \\
                             \bottomrule
    \end{tabular}    
    \begin{tablenotes}
        \item[1] In SSHNN-S, -M, and -L, $F=8$, and $L=4$, $6$, and $8$, respectively.
        \item[2] For a fair comparison, all tests are conducted under semi-supervised learning.
    \end{tablenotes}
    \end{threeparttable}
    \end{adjustbox}
  \end{table}
\vspace{-5mm}
\section{Experimental Results}
\begin{table*}[t]
\vspace{-6mm}
  \caption{Ablation studies on network design and semi-supervised learning. The evaluation metric is Dice and the best results under different fractions are highlighted in bold. Filter multiplier $F=8$.}\label{tableofablation}
  \small
  \centering
  \begin{threeparttable}
  \begin{adjustbox}{scale=0.85}
  \begin{tabular}{cccccccc}
  \toprule\hline
  \multirow{2}{*}{Method} & \multirow{2}{*}{Convolution fusion} & \multirow{1}{*}{U-shaped decoder} & \multicolumn{5}{c}{Unlabeled image count $N_{u}$, labeled image count $N_{l}$ = 1800}            \\ \cline{4-8} 
  &                      & \& Transformer    & 18000 & 5400 & 3600 & 1800 &0 \\   \hline
  \multirow{4}{*}{SSHNN-L} &   {\XSolidBrush}         &     {\XSolidBrush}        & {0.901$\pm$0.001}  & \multicolumn{1}{c}{0.907$\pm$0.001}  & {0.910$\pm$0.003}  & {0.904$\pm$0.003}  & {0.898$\pm$0.005}  \\ 
  &      {\Checkmark}         &      {\XSolidBrush}        & {0.910$\pm$0.001}  & {0.915$\pm$0.001}  & {0.917$\pm$0.002}  & {0.914$\pm$0.003}  & {0.910$\pm$0.004}  \\ 
  &    {\XSolidBrush}          &        {\Checkmark}       & {0.913$\pm$0.001}  & {0.917$\pm$0.002}  & {0.921$\pm$0.002}  & {0.917$\pm$0.002}  & {0.913$\pm$0.004}  \\ 
  &      {\Checkmark}         &      {\Checkmark}      & \textbf{{0.921$\pm$0.000}}  & {\textbf{0.927$\pm$0.002}}  & {\textbf{\underline{0.932$\pm$0.002}}}  & {\textbf{0.929$\pm$0.003}}  &  {\textbf{0.925$\pm$0.003}} \\   \hline
  \multirow{1}{*}{SSHNN-S} &    {\Checkmark}        &    {\Checkmark}   & {0.906$\pm$0.002}  & {0.911$\pm$0.003}  & {0.916$\pm$0.004}  & {0.912$\pm$0.004}  & {0.908$\pm$0.005}  \\ 
  \multirow{1}{*}{SSHNN-M} &    {\Checkmark}        &   {\Checkmark}    & {0.917$\pm$0.001}  & {0.922$\pm$0.001}  & {0.925$\pm$0.002}  & {0.923$\pm$0.003}  & {0.920$\pm$0.004}  \\ \hline
\end{tabular}
\end{adjustbox}
\end{threeparttable}
\vspace{-3mm}
  \end{table*}
\vspace{-3mm}
\subsection{Dataset and Evaluation Metrics}
\vspace{-3mm}
CAMUS \cite{leclerc2019deep} dataset is used to evaluate the performance of SSHNN, which is an open large-scale 
dataset in 2D echocardiography and were collected from 500 patients. Noticed that one four-chamber and one two-chamber view sequences were collected from each patient, lasting around 20 pictures long without labels, except for the moment of end diastole (ED) and end systole (ES). 
Therefore, accessible labelled dataset has 2000 echocardiographic images used for supervised loss calculation, and unlabeled dataset has around 19000 echocardiographic images. Segmentation labels have four types by manual annotations are the left ventricle endocardium, the myocardium, the left atrium and background. In this paper, we adopt the Intersection over Union (IoU), Dice Coefficient (Dice) and Parameters (Params) as the evaluation metrics.
\vspace{-5mm}
\subsection{Implementation Details}
\vspace{-3mm}
We use images of size $256\times 256$ as the network input. Each cell has $B=5$ blocks. For the partial channel connections, $n=4$. Use filter multiplier $F$ and layers $L$ to control the complexity of network and $F$ is initialized as 8. Thus, in $s=4$, there are $B\times F\times \frac{s}{4}=40$ filters. When reducing spatial size from $s\rightarrow 2s$, the number of filters doubles. We use 4 transformers to extract global context. Experiments were carried out on a Nvidia RTX3090Ti GPU. For weights $w$ and architecture $\gamma$, applied optimizer is SGD with momentum 0.9, weight decay 0.0003 and initial value 0.01. For architecture $\alpha$, Adam \cite{kingma2014adam} is applied with learning rate 0.003 and weight decay 0.001. The total number of epochs is 40 and architecture is optimized after 10 epochs since unstable update of $w$ makes local optima. $Teacher$ is updated synchronously with $Student$ for stable training. Finally, use 10\% labeled images in validation and the rest in training with unlabeled images.  
\vspace{-5.5mm}
\subsection{Experimental results}  
\vspace{-3mm}
To demonstrate the effectiveness of SSHNN, we compare it with SOTA approaches. Table \ref{tableofexperiemnts} and Fig .\ref{figure2} respectively elaborate the numerical and visual results for each network on the CAMUS dataset. Mean and standard deviation values for each metric are obtained from cross-validating on the 10 folds of the dataset. From SSHNN series, obviously larger $L$ brings higher model capacity, leading to better segmentation performance at the cost of more parameters (lower speed). Compared with Transfuse, which has suboptimal Dice in Table \ref{tableofexperiemnts}, SSHNN-L gets 0.98\% Dice improvement and 1.05\% IoU enhancement. Similarly compared, SSHNN-M also has the better Dice with fewer parameters, verifying the superiority.
\vspace{-5.5mm}
\subsection{Ablation Study}\label{ablation study}
\vspace{-3mm}
Shown in Table \ref{tableofablation}, we conducted multiple sets of tests to evaluate the impact of convolution fusion, U-shape decoder \& Transformer and the fraction of labeled images. First, SSHNN-L with convolution fusion has a 1.00\% higher Dice than without convolution fusion when unlabeled image versus labeled image $\frac{N_{u}}{N_{l}}=10$. Moreover, use U-shaped decoder with transformer also increases the Dice by 1.33\% compared with no changes at this fraction, verifying the effectiveness of them. Obviously, combine convolution and U-shaped decoder with Transformer works better as SSHNN-L with them remains 0.921 Dice when $\frac{N_{u}}{N_{l}}=10$, aligning with the objective of semi-supervision of learning with minimal supervision. 
\begin{table}[t]
    \vspace{-1mm}
    \caption{Ablation experiments on filter multiplier $F$.}\label{tableofablation2}
  \centering
  \begin{adjustbox}{scale=0.8}
  \begin{tabular}{ccccc}
  \toprule\hline
  \multirow{1}{*}{Model}      & $F$ & Params & Dice$\uparrow$ & IoU$\uparrow$   \\ \hline
    \multirow{3}{*}{SSHNN-L}  & 6   & \textbf{23.6M}  & {0.928$\pm$0.002}    & {0.868$\pm$0.004} \\    
    & 8   &     {38.8M}          & 0.932$\pm$0.002  & {0.873$\pm$0.002}  \\
    & 10 &         {58.2M}   & {\textbf{0.933$\pm$0.001}} & \textbf{{0.876$\pm$0.002}}   \\ 
   \bottomrule
    \end{tabular} 
    \end{adjustbox}
  \end{table}
Note that when $\frac{N_{u}}{N_{l}}=2$, the model outperforms the others as unlabeled data provides extra information and does not mask the characteristics of labeled data. The results of SSHNN-S and SSHNN-M prove the robustness of our method again.

Last but not least, Table \ref{tableofablation2} discusses the effect of $F$, showing that higher model capacity promotes segmentation ability with expensive memory cost, which is a trade-off problem.
\vspace{-5mm}
\section{Conclusion}
\vspace{-4mm}
In this paper, we propose a semi-supervised hybrid NAS network termed SSHNN for medical image segmentation. We adopt convolution fusion to fully utilize multi-scale features extracted by layer-wise searching of HNAS. Furthermore, U-shaped decoder and Transformers are introduced for decode part. Semi-supervised learning Mean Teacher is implemented to overcome limited labeled dataset volumn. Finally, experiments on the CAMUS dataset demonstrate SSHNN realizes a more superior performance than existing methods.

\newpage
\small
\bibliographystyle{IEEEbib}
\bibliography{strings,refs}

\end{document}